\documentclass{elsart}
\usepackage{natbib}
\usepackage{graphicx}

\newcommand {\onde} {O$\nu$DE }

\begin{document}

\runauthor{Giorgio Riccobene}

\begin{frontmatter}

\title{NEMO-\onde: a submarine station for real-time monitoring of acoustic background
installed at 2000 m depth in the Mediterranean Sea}



 \author[INFNCT]{S. Aiello},
 \author[INFNRM]{F. Ameli},
 \author[INFNLNS,UniCT]{I. Amore},
 \author[INFNGE]{M. Anghinolfi},
 \author[INFNLNS]{A. Anzalone},
 \author[INFNNA]{G. Barbarino},
 \author[INFNGE]{M. Battaglieri},
 \author[INFNBA,UniBA]{R. Bellotti},
 \author[INFNBO,UniBO]{M. Bazzotti},
 \author[INFNGE,UniGE]{A. Bersani},
 \author[INFNPI,UniPI]{N. Beverini},
 \author[INFNBO,UniBO]{S. Biagi},
 \author[INFNRM,UniRM]{M. Bonori},
 \author[INFNPI,UniPI]{B. Bouhadef},
 \author[INFNLNS]{G. Cacopardo},
 \author[INFNRM,UniRM]{A. Capone},
 \author[INFNCT]{L. Caponetto},
 \author[INFNBO,UniBO]{G. Carminati},
 \author[INFNBA]{B. Cassano},
 \author[INFNPI,UniPI]{E. Castorina},
 \author[INFNBA]{A. Ceres},
 \author[INFNBO,UniBO]{T. Chiarusi},
 \author[INFNBA]{M. Circella},
 \author[INFNLNS]{R. Cocimano},
 \author[INFNLNS]{R. Coniglione},
 \author[INFNLNF]{M. Cordelli},
 \author[INFNLNS]{L. Cosentino},
 \author[INFNLNS]{M. Costa},
 \author[INFNLNS]{A. D'Amico},
 \author[INFNRM,UniRM]{G. De Bonis},
 \author[INFNBA,UniBA]{C. De Marzo \thanksref{Dead}},
 \author[INFNNA]{G. De Rosa},
 \author[INFNBA]{G. De Ruvo},
 \author[INFNGE]{R. De Vita},
 \author[INFNLNS]{C. Distefano},
 \author[INFNPI,UniPI]{E. Falchini},
 \author[INFNLNS,UniCT]{M. Favetta\thanksref{UniLe}},
 \author[INFNPI,UniPI]{V. Flaminio},
 \author[INFNGE]{K. Fratini},
 \author[INFNBO,UniBO]{A. Gabrielli},
 \author[INFNBO,UniBO]{E. Gandolfi},
 \author[INFNBO,UniBO]{G. Giacomelli},
 \author[INFNBO,UniBO]{F. Giorgi},
 \author[INFNCT]{A. Grimaldi},
 \author[INFNLNF]{R. Habel},
 \author[INFNLNS,UniCT]{G. Larosa\thanksref{UPV}},
 \author[INFNCT,UniCT]{E. Leonora},
 \author[INFNRM]{A. Lonardo},
 \author[INFNNA,UniNA]{G. Longo},
 \author[INFNCT,UniCT]{D. Lo Presti},
 \author[INFNRM,UniRM]{F. Lucarelli},
 \author[INFNPI,UniPI]{E. Marinelli},
 \author[INFNBO,UniBO]{A. Margiotta},
 \author[INFNLNF]{A. Martini},
 \author[INFNRM,UniRM]{R. Masullo},
 \author[INFNBA,UniBA]{R. Megna},
 \author[INFNLNS,UniCT]{E. Migneco},
 \author[INFNGE]{S. Minutoli},
 \author[INFNBA]{M. Mongelli},
 \author[INFNPI,UniPI]{M. Morganti},
 \author[INFNGE]{P. Musico},
 \author[INFNLNS]{M. Musumeci},
 \author[INFNLNS]{A. Orlando},
 \author[INFNGE]{M. Osipenko},
 \author[INFNNA]{G. Osteria},
 \author[INFNLNS]{R. Papaleo},
 \author[INFNLNS]{V. Pappalardo},
 \author[UniPV,INFNLNS]{G. Pavan},
 \author[INFNCT,UniCT]{C. Petta},
 \author[INFNLNS]{P. Piattelli},
 \author[INFNGE]{D. Piombo},
 \author[INFNLNS,UniCTDMFCI]{S. Privitera\thanksref{nomore}},
 \author[INFNLNS]{G. Raia},
 \author[INFNCT]{N. Randazzo},
 \author[INFNCT]{S. Reito},
 \author[INFNGE,UniGE]{G. Ricco},
 \author[INFNLNS]{G. Riccobene\corauthref{ca:fax}}
 \ead{riccobene@lns.infn.it},
 \author[INFNGE]{M. Ripani},
 \author[INFNLNS,UniCT]{D.J. Romeo\thanksref{nomore}},
 \author[INFNLNS]{A. Rovelli},
 \author[INFNBA,UniBA]{M. Ruppi},
 \author[INFNCT,UniCT]{G.V. Russo},
 \author[INFNNA,UniNA]{S. Russo},
 \author[INFNLNS]{P. Sapienza},
 \author[INFNLNS]{M. Sedita},
 \author[MSU]{E. Shirokov},
 \author[INFNRM,UniRM]{F. Simeone},
 \author[INFNCT,UniCT]{V. Sipala},
 \author[INFNLNS]{F. Speziale},
 \author[INFNBO,UniBO]{M. Spurio},
 \author[INFNGE,UniGE]{M. Taiuti},
 \author[INFNPI]{G. Terreni},
 \author[INFNLNF]{L. Trasatti},
 \author[INFNCT]{S. Urso},
 \author[INFNLNF]{V. Valente},
 \author[INFNRM,UniRM]{M. Vecchi},
 \author[INFNRM]{P. Vicini},
 \author[INFNRM,UniRM]{R. Wischnewski}.


\corauth[ca:fax]{Fax: +39 095 542 398}
\thanks[Dead]{Deceased}
\thanks[UniLe]{Present address, Dipartimento di Fisica Universit\`a del Salento, Via Arnesano, 73100, Lecce, Italy }
\thanks[UPV]{Present address, Universidad Politecnica de Valencia, Campus de Gandia, Ctra. Nazaret-Oliva, 46730, Grao de Gandia, Valencia, Spain}
\thanks[UniCTDMFCI]{Present address, Dipartimento di Metodologie Fisiche e Chimiche per l'Ingegneria Universit\`a di Catania, Viale A. Doria 6, 95125 Catania, Italy }
\thanks[nomore]{Presently in private occupation}

\address[INFNLNS]{Laboratori Nazionali del Sud INFN, Via S.Sofia 62, 95123, Catania, Italy}
\address[INFNLNF]{Laboratori Nazionali di Frascati INFN, Via Enrico Fermi 40, 00044, Frascati (RM), Italy}
\address[INFNBA]{INFN Sezione Bari, Via Amendola 173, 70126, Bari, Italy}
\address[INFNBO]{INFN Sezione Bologna, V.le Berti Pichat 6-2, 40127, Bologna, Italy}
\address[INFNCT]{INFN Sezione Catania, Via S.Sofia 64, 95123, Catania, Italy}
\address[INFNGE]{INFN Sezione Genova, Via Dodecaneso 33, 16146, Genova, Italy}
\address[INFNNA]{INFN Sezione Napoli, Via Cintia, 80126, Napoli, Italy}
\address[INFNPI]{INFN Sezione Pisa, Polo Fibonacci, Largo Bruno Pontecorvo 3, 56127, Pisa, Italy}
\address[INFNRM]{INFN Sezione Roma 1, P.le A. Moro 2, 00185, Roma, Italy}

\address[UniBA]{Dipartimento Interateneo di Fisica Universit\`a di Bari, Via Amendola 173, 70126, Bari, Italy}
\address[UniBO]{Dipartimento di Fisica Universit\`a di Bologna, V.le Berti Pichat 6-2, 40127, Bologna, Italy}
\address[UniCT]{Dipartimento di Fisica e Astronomia Universit\`a di Catania, Via S.Sofia 64, 95123, Catania, Italy}
\address[UniGE]{Dipartimento di Fisica Universit\`a di Genova, Via Dodecaneso 33, 16146, Genova, Italy}
\address[UniNA]{Dipartimento di Scienze Fisiche Universit\`a di Napoli, Via Cintia, 80126, Napoli, Italy}
\address[UniPI]{Dipartimento di Fisica Universit\`a di Pisa, Polo Fibonacci, Largo Bruno Pontecorvo 3, 56127, Pisa, Italy}
\address[UniPV]{Centro Interdisciplinare di Bioacustica e Ricerche Ambientali, Dipartimento di Biologia Animale Universit\`a di Pavia, Via Taramelli 24, 27100, Pavia, Italy}
\address[UniRM]{Dipartimento di Fisica Universit\`a La Sapienza, P.le A. Moro 2, 00185, Roma, Italy}
\address[MSU]{Faculty of Physics, Moscow State University, 119992, Moscow, Russia}

\begin{abstract}

The NEMO (NEutrino Mediterranean Observatory) Collaboration
installed, 25 km E offshore the port of Catania (Sicily) at 2000 m
depth, an underwater laboratory to perform long-term tests of
prototypes and new technologies for an underwater high energy
neutrino km$^3$-scale detector in the Mediterranean Sea. In this
framework the collaboration deployed and successfully operated for
about two years, starting form January 2005, an experimental
apparatus for on-line monitoring of deep-sea noise. The station
was equipped with 4 hydrophones and it is operational in the range
30 Hz - 43 kHz. This interval of frequencies matches the range
suitable for the proposed acoustic detection technique of high
energy neutrinos. Hydrophone signals were digitized underwater at
96 kHz sampling frequency and 24 bits resolution. A custom
software was developed to record data on high resolution
4-channels digital audio file. This paper deals with the data
analysis procedure and first results on the determination of sea
noise sound pressure density curves. The stored data library,
consisting of more than 2000 hours of recordings, is a unique tool
to model underwater acoustic noise at large depth, to characterise
its variations as a function of environmental parameters,
biological sources and human activities (ship traffic, ...), and
to determine the presence of cetaceans in the area. \vspace{1pc}
\end{abstract}


\end{frontmatter}

\section{Introduction}

In recent years the astrophysics and particle physics Community
strongly addressed its efforts in the realization of large
experimental apparatuses with the goal of detecting high energy
neutrinos ($E_\nu > 10^{12}$ eV) originated in cosmic sources
\cite{GaisserHalzenStanev1995}. The detection of these particles
will open a new window in our comprehension of the Universe and in
the understanding of physics processes occurring in powerful
cosmic sources such as Active Galactic Nuclei \cite{Mannheim1995},
Gamma Ray Busters \cite{WaxmanBahcall1997}, microquasars
\cite{Distefano2002} and Supernova Remnants \cite{Vissani2005}.

The experimental techniques proposed to identify the cosmic
neutrino signatures are mainly three \cite{Markov1961}: the
detection of \v{C}erenkov blue light originated by charged leptons
(electrons, positrons, muons and tauons) from a neutrino
interaction in water or ice; the detection of acoustic waves
produced by neutrino energy deposition in water, ice or salt; the
detection of radio pulses following a neutrino interaction in ice
or salt.

Due to the faintness of expected astrophysical neutrino fluxes
(about 10$^6$ particles per km$^2$ per year, following a $E^{-2}$
spectrum), and due to the extremely low probability of neutrino
detection in the detector volume ($\simeq 10^{-4}$), the required
size of these apparatuses is about 1 km$^3$ for neutrinos of
energy $10^{12}\div10^{15}$ eV and more than 10 km$^3$ for more
energetic particles. The only way to build such a detector is the
use of large natural media such as seawater, polar ice or lake
freshwater. Moreover all these techniques require that the
detector must be placed at large depth underground (about 1000 m),
undersea or under ice ($>$3000 m) to shield the detector from the
intense background of cosmic particles that would cancel the
neutrino signal. The \v{C}erenkov technique is presently in a
mature phase. The first km$^3$ scale detector is under
construction at the South Pole \cite{Icecube}.The activity in the
Mediterranean Sea in presently focused in the construction and
operation of prototypal, small scale detectors: ANTARES
\cite{Carr2006}, NEMO \cite{Migneco2006} and NESTOR
\cite{Resvanis2006}, technological demonstrators for the future
Km3Net \cite{Km3Net}, whose construction is planned between 2009
and 2012.

The NEMO Collaboration is strongly involved in the design and
construction of the Mediterranean km$^3$ \v{C}erenkov neutrino
detector. In the same time the Collaboration is starting studies
on the acoustic detection technique; the first task, in this
framework, was the measurement and monitoring of the acoustic
background at large depth to evaluate the expected signal to noise
ratio. At present, only few measurements of acoustic noise have
been carried out at very large depth, where acoustic detectors
should be presumably located. This is mainly due to technological
difficulties in constructing, deploying and operating real-time
monitoring stations in deep sea. Noise in the sea has different
origins: biological (fishes, marine mammals, crustaceans), seismic
and micro-seismic, mechanical (wind and surface waves), molecular
thermal vibrations and human activities (navigation, fishing,
military operations, oceanographical instrumentation, oil
exploration). Biological and human noises could reach very high
pressure level, but they are, generally, produced by local and
intermittent sources. At large depths it is expected that surface
agitation noise (which is the major source of noise in the kHz
range) should be reduced due to the change of sound refraction
index with depth. On the other hand, it is not well known the
contribution of sound emissions generated by cetaceans, that can
immerse down to thousand meters depth \cite{Pavan2000}.
Bibliographic data indicate that, in the frequency range of
interest for neutrino detection (10 $\div$ 100 kHz), the acoustic
noise in water is a sum of a diffuse and relatively steady
background due to ship traffic and sea state conditions that
occasionally add up with loud and transient sources, such as
biological sounds (dolphin and whale vocalizations), and man-made
noise (close ships, navigation and scientific instrumentation:
pingers, air-guns) \cite{Urick}.

In order to measure the level of acoustic noise in the deep
Mediterranean Sea, the NEMO Collaboration constructed and operated
the experimental station O$\nu$DE (Ocean noise Detection
Experiment), a real-time experiment to monitor acoustic signals in
deep sea.

\section{The physics case}

Underwater \v{C}erenkov telescopes for high energy neutrinos are
arrays of large-area photomultipliers (PMTs), having typically 10
inches diametre, designed to detect \v{C}erenkov blue light
radiated in water by charged leptons from neutrino Charged Current
interactions. The reconstruction of the \v{C}erenkov tracks allows
the identification of the lepton direction and energy thus, to
some extent, the neutrino direction and energy. The effort of the
astro-particle physics community is presently addressed in the
construction of two detectors (Km3Net \cite{Katz2006} in the
Mediterranean Sea and ICECUBE \cite{Achterberg2007} in the
Antarctic ice-cap) equipped with some thousands of PMTs. These
detectors are expected to reach detection areas of about 1 km$^2$
at E$_\nu =$ 1$\div$100 TeV and to identify astrophysical
point-like neutrino sources and measure the high energy diffuse
cosmic neutrino flux. At higher energies ($>10^{16}$ eV) however,
the expected neutrino fluxes are fainter (the spectrum follows a
$E_\nu^{-2}$ power law), and detectors with detection areas $>$10
km$^2$ are required. A different detection technique was suggested
\cite{Askarian1957} to build larger detectors. At these energies
neutrino interactions in water produce showers, that release
instantaneously a macroscopic amount of energy in a small
cylindrical volume of matter. Ionization and sudden heating of
water produce a bipolar pressure pulse which expands
perpendicularly to the shower axis. The maximum wave amplitude,
calculated with thermo-acoustic models, scales linearly with the
density of energy deposition: for a 10$^{20}$ eV neutrino induced
shower it is few tens mPa at a distance of $\sim$ 1 km from the
shower axis \cite{Vandenbroucke2005}. The wave peak frequency is
estimated to be in the range of 10 kHz. The advantage of acoustic
detection is the long absorption length of sound in water: in this
frequency range, it is of the order of few kilometres. A
pioneering work on acoustic neutrino detection has been recently
conducted using military arrays of hydrophones
\cite{Lehtinen2002}.

\begin{figure}[h]
\vspace*{-2.0mm}
\centerline{\includegraphics[width=14cm]{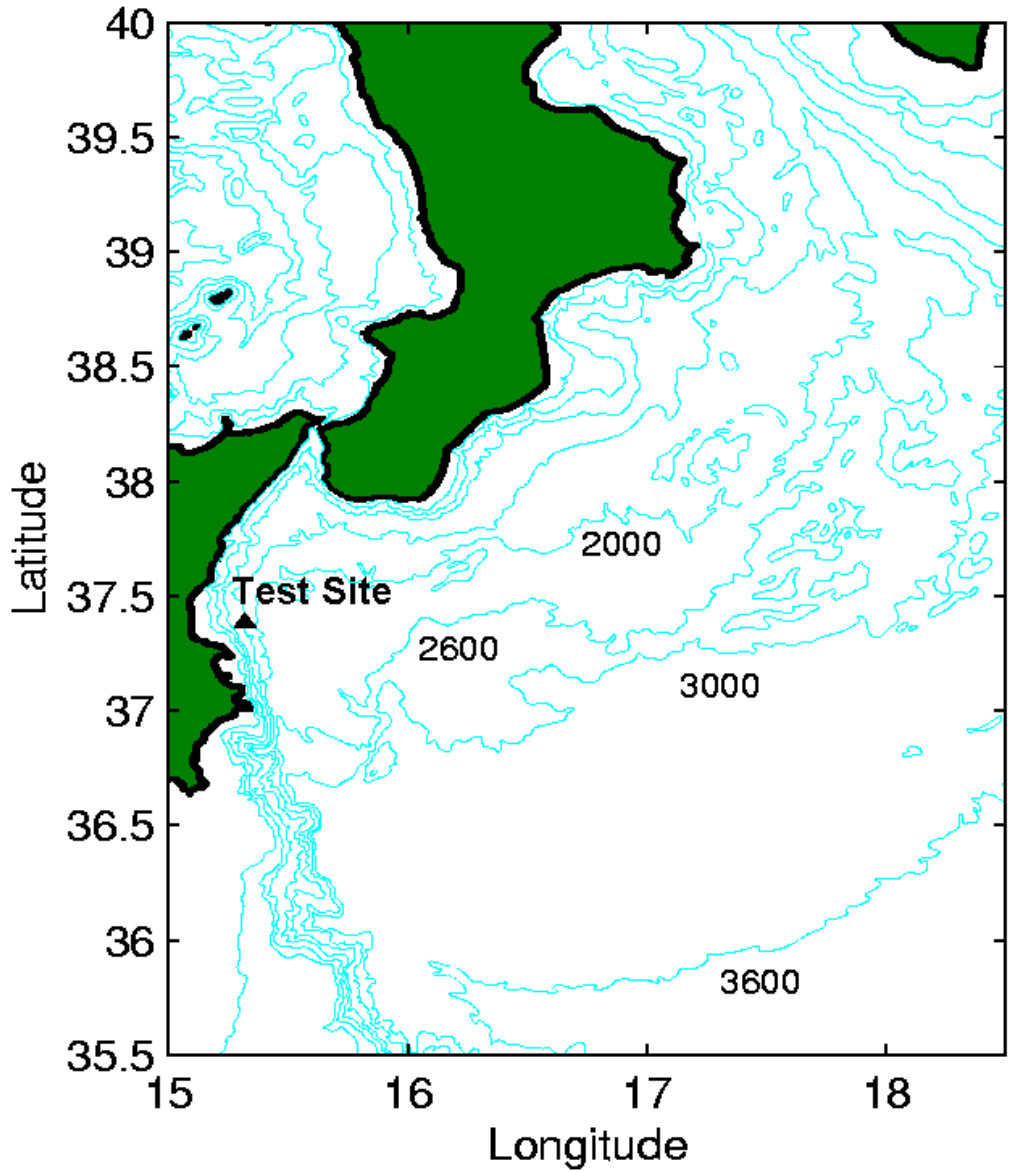}}
\caption{ Bathymetric chart of the Eastern Sicilian Coast region.
The location of the Catania TSS (triangle) is shown. The \onde
frame is moored at 2050 m depth, latitude $37^{\circ}30'008$ N and
longitude $015^{\circ}23'034$ E. \label{fig:fig1_Site}}
\end{figure}

Due to the small amplitude of the expected neutrino bipolar
signal, it is mandatory to measure the acoustic noise in the sea
as a function of frequency in order to study the performances of a
future acoustic detector as a function of the number of sensors
and of the design of the antenna. This was the main goal of \onde.
The detector was deployed during January 2005 at the INFN
Laboratori Nazionali del Sud (LNS) deep sea Test Site, located at
depth of $\sim$ 2000 m, 25 km E offshore the port of Catania
(Sicily), see Figure  \ref{fig:fig1_Site}. The detector acquired
data from January 2005 to November 2006, when the NEMO
Collaboration started to install the NEMO Phase 1 detector: a
technological demonstrator for the future km$^3$ \v{C}erenkov
neutrino telescope \cite{Migneco2006}.

\section{The Catania Test Site infrastructure}

\begin{figure}[h]
\centerline{\includegraphics[width=14cm]{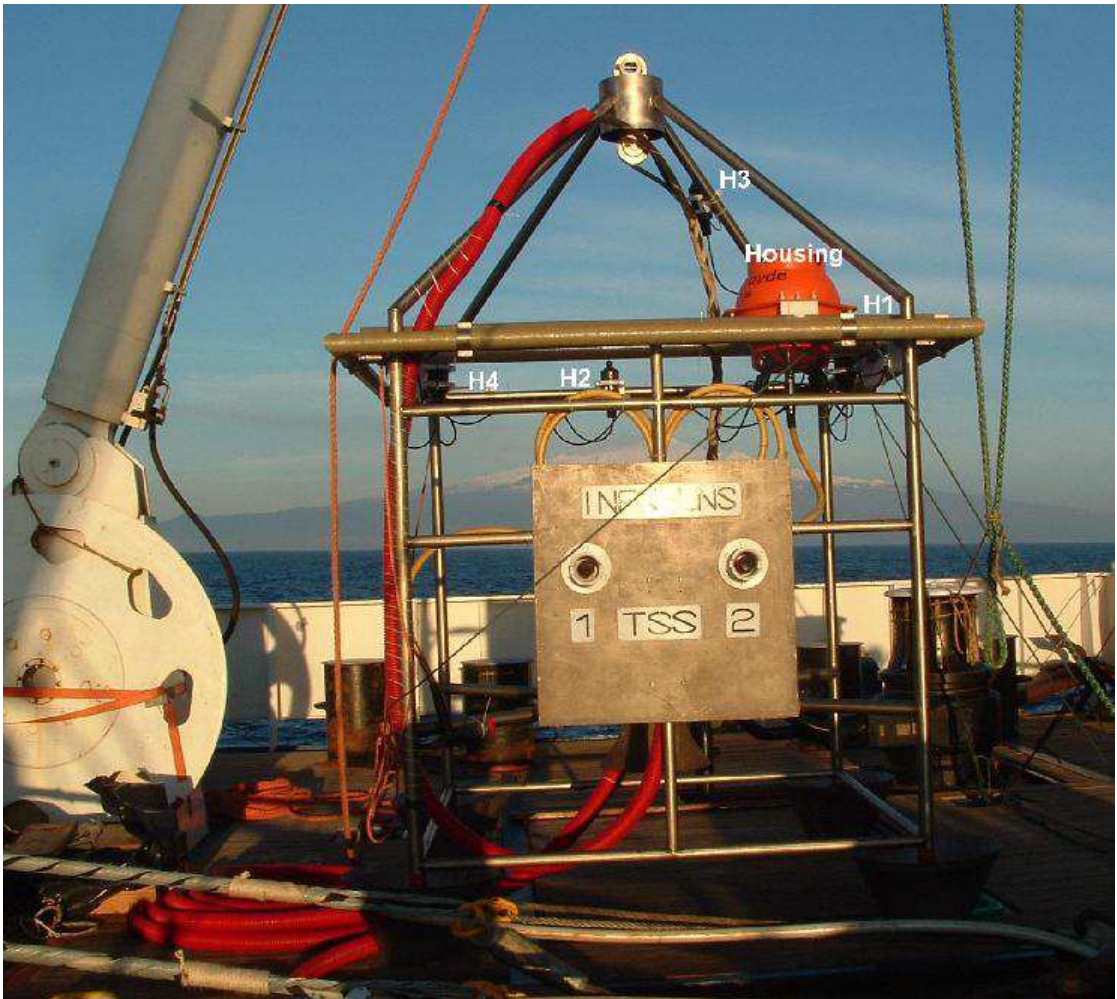}}
\caption{The titanium frame installed on TSS. The ROV operable
electro-optical connectors are visible on the front panel. The
hydrophones and electronics housing of the O$\nu$DE station are
also shown (see text).} \label{fig:fig2_frame}
\end{figure}

The Catania Test Site consists of a shore laboratory, a 28 km long
electro-optical (hereafter e.o.) cable connecting the shore lab to
the deep sea lab. The shore laboratory hosts the land termination
of the cable, the on-shore data acquisition system and power
supplies for underwater instrumentation. The underwater cable is
an umbilical underwater e.o. cable, armored with an external steel
wired layer, containing 10 optical single-mode fibers (standard
ITU-T G-652) and 6 electrical conductors (4 mm$^2$ area). At about
20 km E from the shore, the cable is divided into two branches,
roughly 5 km long each, that reach two different sites namely Test
Site North (latitude $37^{\circ}30'810$ N, longitude
$015^{\circ}06'819$ E depth 2100 m), and Test Site South (latitude
$37^{\circ}30'008$ N, longitude $015^{\circ}23'034$ E, depth 2050
m). The Test Site North (TSN) cable branch has 2 conductors and 4
fibres directly connected to shore, the Test Site South (TSS)
branch has 4 conductors and 6 fibers. After deploying the main
underwater cable, in January 2005 the Collaboration installed, on
TSS and on TSS, two underwater frames. Each frame, made of grade 2
titanium, is equipped with a pair of e.o. connectors (see figure
\ref{fig:fig2_frame}). The two frames were deployed on the seabed.
The e.o. connectors are made to be handled by underwater robots
ROV (Remotely Operated Vehicles) to allow plugging and unplugging
of underwater experimental apparatuses, avoiding further recovery
operations of the main cable. During the same naval campaign two
experimental apparatuses were deployed, plugged and put in
operation. The seismic and environmental monitoring station
Submarine Network 1 (SN-1), designed and operated by the INGV
(Istituto Nazionale di Geofisica e Vulcanologia, Italy)
\cite{Favali2005,Favali2006} was connected to the TSN termination.
This station is presently the only cabled node of the ESONET
(European Seafloor Observatory NETwork) project \cite{ESONET}. In
January 23$^{th}$ 2005 the \onde station was deployed and
connected to the TSS termination.

\section{The O$\nu$DE apparatus}

\onde was designed to perform on-line monitoring of the acoustic
noise at large depth. The station is equipped with four large
bandwidth hydrophones (30 Hz - 50 kHz). Each hydrophone (hereafter
H1,H2 H3 and H4) was mounted on an aluminum alloy vessel, pressure
resistant, which also hosted the hydrophone preamplifier. The
analog signals from the preamplifiers were transmitted, through
underwater cables suitable for audio applications, to signal
conditioning and digitization electronics hosted in a
pressure-proof glass housing. Underwater, digital signals were
translated into optical and sent to shore through the optical
fibers. On shore, acoustic data were reconverted into electrical
and recorded using a PC, in which a pair of professional PCI audio
boards were mounted. Electrical power was supplied from shore.

\subsection{Mechanical set-up}

The mechanical structure of \onde is composed by: a commercial
pressure-proof glass housing (which hosts the DAQ and power supply
electronics), one electro optical cable that connects the station
to the e.o. plug mounted on the frame and four electrical cables
that connect the housing to the four hydrophone vessels.

\begin{figure}[h]
\centerline{\includegraphics[width=14cm]{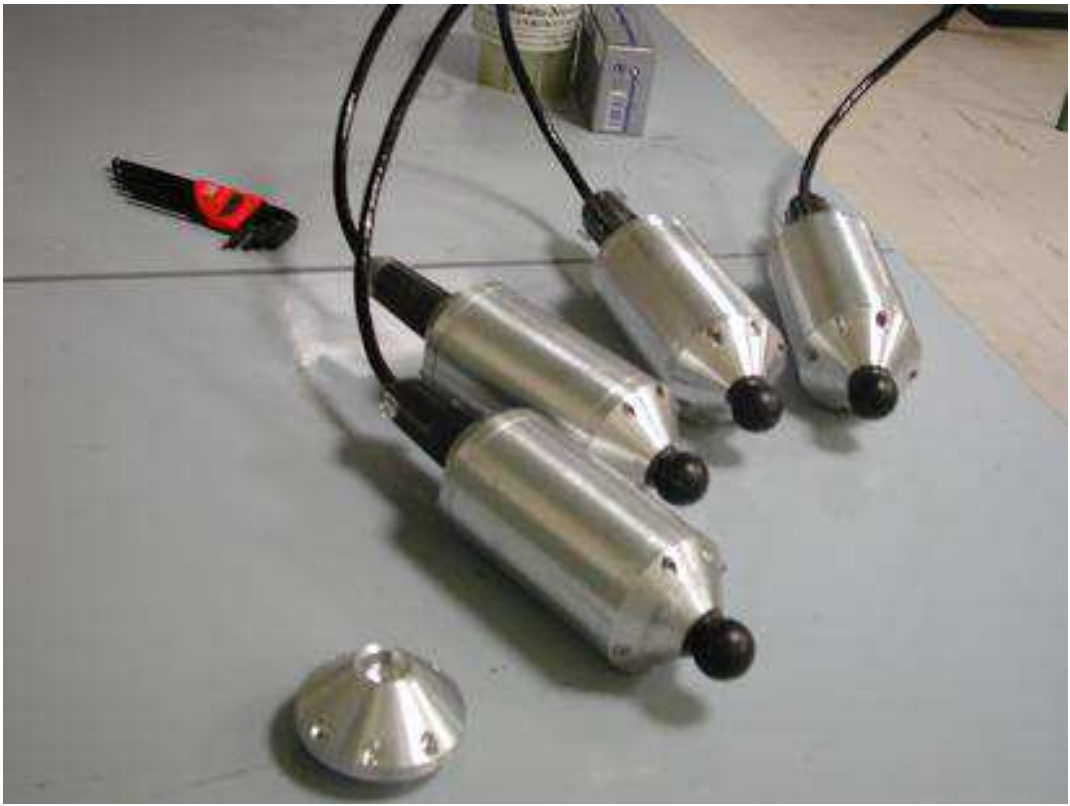}}
\caption{The Al-alloy housing that hosts preamplifiers and holds
the hydrophones. \label {fig:fig3_hydros}}
\end{figure}

The vessels containing the hydrophone preamplifiers were made in
aluminum alloy (Al-7075). The vessels are cylindrical with
trunk-conical shaped terminations (angle 45$^{\circ}$). This shape
was chosen to fit the hydrophone body and to minimize the
reflections of acoustic waves towards the hydrophone. A penetrator
at the base of the aluminum vessel was designed and constructed to
insert the electrical cable that connects the vessel with the
glass housing (see Figure \ref{fig:fig3_hydros}).

The hydrophones vessels were hooked on the upper part of the TSS
frame, forming a tetrahedral antenna of $\sim$ 1 m side. The
hydrophone vessel H3, was mounted in the highest position, close
to the frame apex, that is at about 3.2 m above the seabed. Since
most of the noise comes from above (the station is moored on the
seabed) H3 was used as pilot hydrophone during signal analysis.
H1,H2 and H4 were attached approximately at the same height (about
2.6 m above the seabed), in the squared upper edge of the frame.
In picture \ref{fig:fig2_frame}, H1 is visible on the right with
respect to the instrument housing (the orange spherical shell), H2
is placed behind the shell and H4 on the left. The glass housing
is a commercial 17'' diametre sphere, manufactured by {\em
Nautilus} \cite{Nautilus}. The sphere is made of two halves: the
electronics was placed inside the sphere and, before deployment,
the two halves were sealed together slightly de-pressuring to 750
mbar the cavity of the sphere, in a nitrogen filled environment.
This pressure ensures both the sealing and the circulation of
nitrogen inside the sphere, thus the cooling of the electronics.
The sphere was equipped with 5 titanium connectors. One is an
electro-optical dry-mateable connector, holding 3 optical and 2
electrical contacts. This connector was used to link the station
to the frame e.o. connectors, by means of an electro-optical
harness cable. The harness was terminated on one side with a dry
mateable-plug, on the other with a ROV-mateable connector that
matches the one installed in the frame. The other four electrical
connectors were used to link independently each hydrophone to the
electronics glass housing.

In order to get the absolute orientation of the frame (thus the
orientation of the station), and in order to monitor the possible
movements of the frame with respect to seabed, a  tilt-meter and
compass board (EZ-Compass-3, manufactured by {\it Advanced
Orientation System} \cite{AOSI}) was installed inside the
electronics glass housing of \onde. The compass was equipped with
an RS-232 communication port, which allowed board initialization,
calibration and serial data transmission. The compass board was
connected to shore through an optical bi-directional WDM optical
link. The compass indicated that the frame had an orientation of
about 108 $\pm$2 degrees with respect to North and it was laid
almost co-planar ($5^{\circ}\pm 2^{\circ}$ roll,
$-3^{\circ}\pm2^{\circ}$ pitch) with the seabed. These values
where confirmed by ROV inspections after the deployment and have
not changed during the whole period of data taking.

\subsection{Hydrophones and pre-amplifiers}

We used {\em RESON} \cite{Reson} TC-4042C hydrophones, derived
from the TC4037 Series and tested by the manufacturer to operate
at 250 bar pressure for long term deployment. The used hydrophones
are piezoelectric sensors, having a mean receiving sensitivity of
-195$\pm$3 dB re 1V/$\mu$Pa, linear over a wide range of
frequencies: from few tens Hz to about 50 kHz\footnote{We remind
to the reader that the hydrophone sensitivity is defined as the
sensor transduction factor V over $\mu$Pa, thus it is not the
minimum value of pressure detectable by the sensor. The used
hydrophones have a sensitivity of -195 dB thus they convert an
acoustic signal of 1$\mu$Pa into an electric signal of $\sim$ 1.78
nV}. The hydrophone analog output is differential. The TC-4042C
hydrophones were mounted on the channels H1,H2 and H3. A
hydrophone from a different series, having a sensitivity 5 dB
lower, was mounted on channel H4. All hydrophones have an
omnidirectional directivity pattern suitable for ambient noise
measurements, which is the purpose of the experiment. The
hydrophone output signal was feeded into a preamplifier, developed
also by {\em RESON}, which has a gain of 20 dB. Two preamplifiers
(namely the ones installed on channels  H2 and H4) were modified
applying a hi-pass filter ($>$ 1 kHz, 6 dB per octave) to reduce
the expected ambient noise, which has typically $1/f$ spectrum.
This was done to avoid possible saturations due to the low
frequency noise and to focus the measurements to the frequency
range interesting for neutrino detection ($>$10 kHz). On the other
hand the use of a pair of unfiltered large-bandwidth hydrophones
(H1 and H3) allowed comparison with bibliographic data, which is
more abundant for low frequency measurements.

\subsection{Data digitization and transmission electronics}
\label{sec:electronics}

The differential output of each preamplifier was sent to a pair of
line-output and line-input transformers. Line transformers were
used to galvanically insulate the lines in case of shorts inside
the hydrophone vessels and to balance the audio line. The
line-output transformer was hosted inside the aluminum vessel, the
line-input transformer inside the glass housing. The electrical
line between the transformers, 4 m long, was a shielded
twisted-pair cable suitable for analogue audio signal
transmission.

The hydrophones signals were then sent to two stereo Analog to
Digital Converters (ADC). Signal digitization was performed using
{\it Crystal} CS5396 stereo ADCs \cite{Crystal}. In particular
channel H1 and H3 were plugged in the left and right channel of
one board respectively, H2 and H4 (modified applying a $f>$1 kHz
hi-pass filter, 3 dB per decade) were plugged to the left and
right channels of the other board. The two ADCs received the same
12.288 MHz clock, thus they were synchronised. The CS5396 is a
sigma delta ADC which samples the analog data at a rate of 96 kHz
with a resolution of 24 bits, the input voltage range of the ADC
is 4 V$_{\hbox{PP}}$. The ADC outputs were sent to a Digital
Interface Transmitters {\it Crystal} CS8404A that converted the
data stream into standard SPDIF (Sony Philips Digital Interface
Format) stream. The SPDIF protocol contains, together with data,
the sampling time information; since the two ADCs and the two
digital audio transmitters were driven by the same common clock
the two stereo streams are synchronized. Since we know the phase
response of the
 $>1$ kHz hi-pass filters applied on H2 and H4, the whole array  can
be also phased. This feature is extremely useful for TDoA (Time
difference of arrival) analysis of signals detected by the four
hydrophones, in order to recover the direction of emission of the
detected sounds. The two output streams were sent to a pair of
electro-optical media converters capable to transmit data over
$\sim$ 50 km single mode optical fibre.

\subsection{Power Supply}

Power was supplied to the station from shore using stabilized 220
V$_{\hbox{ac}}$ 50 Hz supplier and it was elevated in order to
feed the station constantly at about 380 V$_{\hbox{ac}}$. Taking
into account the cable impedance, the voltage at shore was set, by
means of a VARIAC, to about 415 V$_{\hbox{ac}}$. A zero crossing
switch was also used on shore to start power transmission only
when voltage sinusoid was rising and close to zero. The power
supply system was realised using all linear components, avoiding
switching power converters which could cause electrical and
mechanical noise in the interesting frequency band. Moreover all
electronic components were galvanically insulated one from the
other to avoid propagation of short-circuits through the power
chain. Inside the deep sea housing we placed independents ac/ac
power transformers, rectifiers and regulators to supply
electronics at different dc currents (14 V$_{\hbox{dc}}$ and 5
V$_{\hbox{dc}}$), and to separate the power lines of analogue
electronics from the digital ones. The underwater power
distribution was made using low drop-out regulators to minimize
power dissipation: the transformation efficiency was $\sim 70\%$
and the total power consumption, at 380 V$_{\hbox{ac}}$ supply,
was about 20 W. The temperature recorded close to the ADCs, by a
thermometer mounted on compass board, was during the whole
operation period, 26.2$\pm$0.2 $^{\circ}$C, when the sea
temperature was 13.8$\pm$0.1 $^{\circ}$C.

\section{On shore data acquisition}

On shore, data from the underwater station were re-translated into
electrical audio SPDIF (Sony Philips Digital Interface Format)
standard using a pair of fiber optical data receivers. The two
SPDIF stereo data stream are then addressed to a PC (Pentium IV, 3
GHz, 1 GB RAM) equipped with two professional PCI audio boards,
RME DIGI96-8 PAD \cite{RME}. In this sections the data
acquisition/recording software and the file archiving strategy are
described.

\subsection{Software tools}

Data acquisition on shore was performed, from January to April
2005, using standard 16 bits audio software tools and sampling
independently the two pairs of hydrophones (H1-H3 and H2-H4). From
May 2005 we used a custom software tool {\it SeaRecorder}
developed by CIBRA \cite{CIBRA} running under Windows XP. The
program reads and keeps synchronized the two digital stereo data
streams coming from the underwater station. Data, received in
SPDIF format at 24 bit resolution and 96 kHz sampling frequency,
are saved into standard Microsoft {\tt .wav} 32 bit float format
(24+8 bit).  This format was chosen to allow data porting to
Matlab \cite{matlab} for off-line analysis.

\begin{figure}[h]
\centerline{\includegraphics[width=14cm]{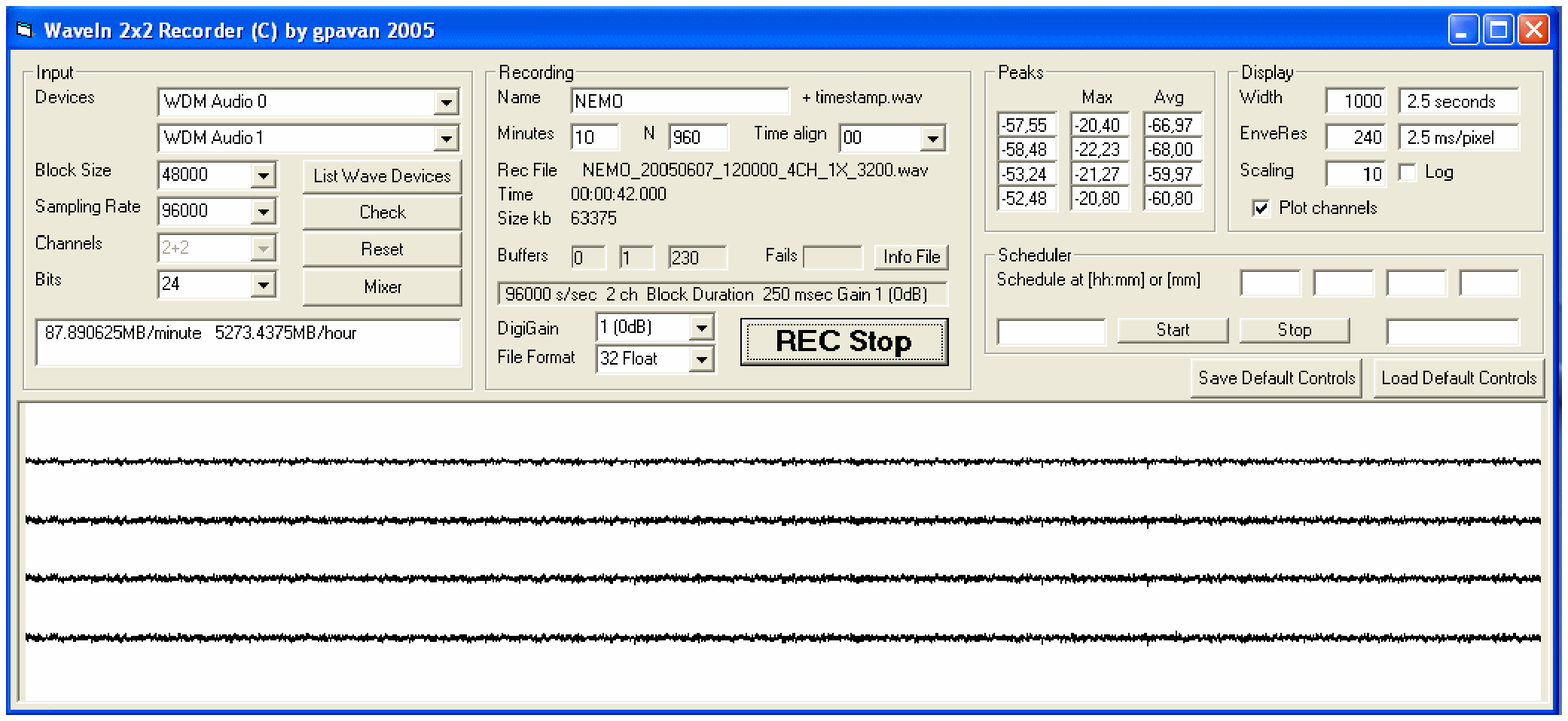}}
\caption{The main window of the SeaRecorder program, used to
record and monitor data coming from the four hydrophones installed
on \onde \label {fig:fig4_SeaRec}}
\end{figure}

{\it SeaRecorder} permits both data recording with floating point
format or integer (16 or 32 bit/sample), and digital amplification
of data at several gain factors. During acquisition, the software
displayed average and peak values measured by the 4 channels and
plotted, in real time, their envelope to provide on-line
monitoring of the recording. The program also generated a log text
file containing complete information of the software settings,
average and maxima values measured for each recording. In figure
\ref{fig:fig4_SeaRec} the data acquisition window of the used
software is shown.

File recording can be programmed to be continuous, with automatic
file splitting every hour or every 30 minutes, or scheduled for
predefined file duration. A special filenaming protocol was
adopted to reduce the risk of data losses or data
misinterpretation. Filenames included a date-and-time stamp,
number of channels, recording gain (linear), file format; sample
rate was omitted as it was a hardware-locked parameter (96 kHz); a
typical filename was therefore {\tt
ONDE\_20050827\_161500\_4CH\-1X\_3200.wav}.

An additional software tool, the {\it SeaPro} also developed by
CIBRA, was used to read off-line the 4 channels, 32 bit, 96 kHz
{\tt .wav} files recorded by {\it SeaRecorder} to permit detailed
data visualization. Two channels (selectable among the four) can
be played as sounds and visualized as spectrograms (time vs.
frequency diagrams) in the same time, allowing identification and
classification of different sounds.

\subsection{Data archival}

After the experiment start-up, the data were continuously recorded
for about one month, this allowed to evaluate the average value
and variability of sound level and to define the successive
strategy for scheduled recording. Continuous recording strategy
was not possible due to storage space constraints: the amount of
data sent to shore requires about 124 GB/day. Data were therefore
recorded for five minutes (randomly chosen) continuatively every
hour, this was a compromise to save a representative sample of
unbiased data, reducing disk space consumption: the storage space
required daily for 4 channel recording at 32 bits was 10.2 GB. A
larger sample of data (about 20' per hour) coming from H3 only,
was also recorded using 16 bit file format.

The data sample presently analysed amounts to $\sim$ 1200 hours,
covering 16 months from January to December 2005, and from July to
November 2006. From mid February to the end of March 2005, and
from January to June 2006, the station was not in operation due
maintenance of the e.o. main cable and on-shore hardware. As
explained in the following, the present paper deals with data
recorded from May 2005 on.

\section{Data analysis}

As previously described, data from the four hydrophones were
recorded as 4 channels {\tt .wav} files at 24 bits and 96 kHz.
This permitted offline data analysis  under Matlab environment.

\begin{figure}[h]
\centerline{\includegraphics[width=14cm]{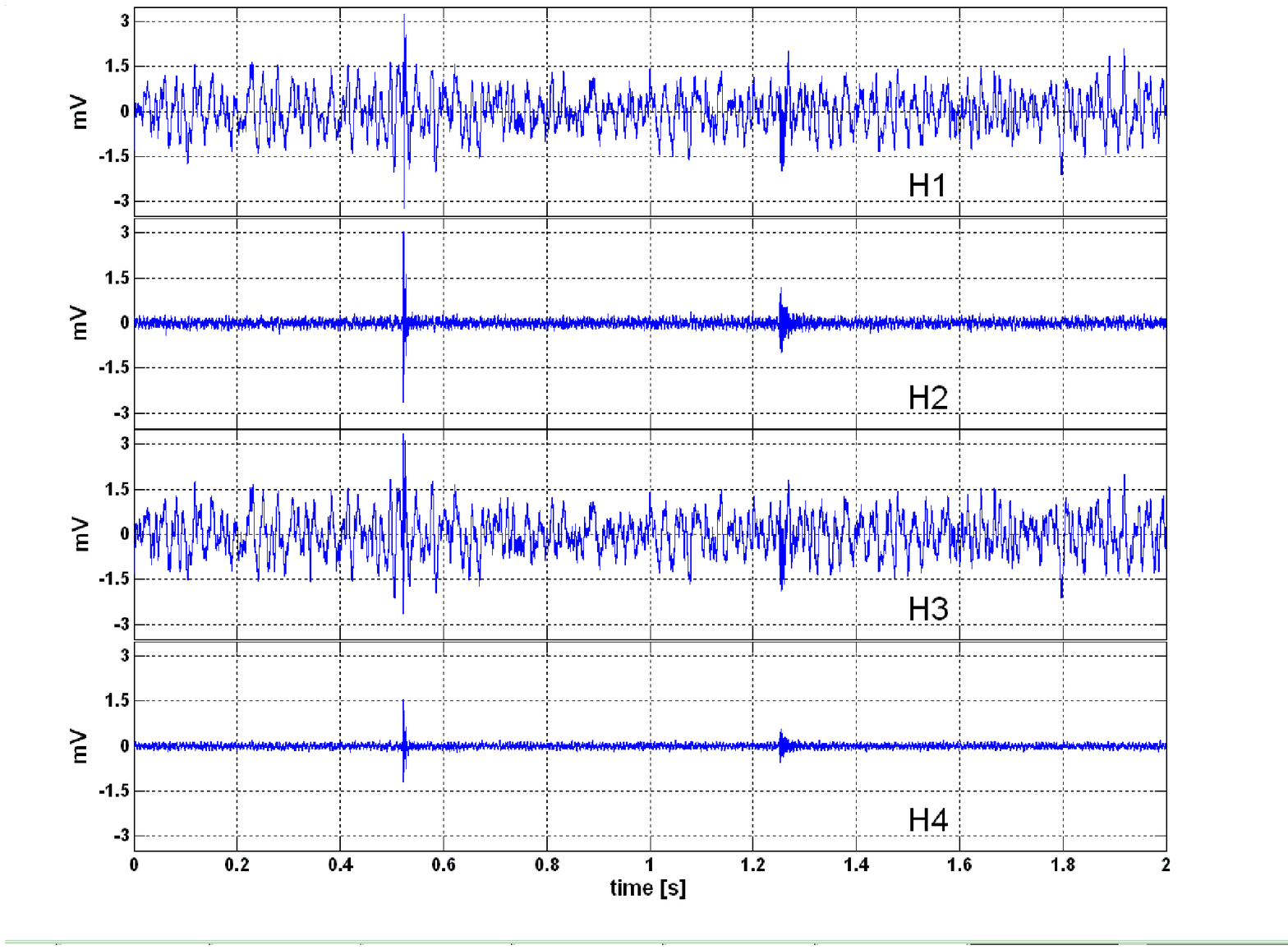}}
\caption{\label{fig:fig5_long_click}  Example of recorded raw data
(only 50 Hz noise filtered): a sperm whale click (occurring at
$\sim$ 0.5 s) and its reflection on the sea surface (occurring at
$\sim 1.3$ s). The four hydrophone channels are independently
displayed. H2 and H4 had a hardware hipass filter $f>$1kHz and the
whale click is clearly visible.}
\end{figure}

In figure \ref{fig:fig5_long_click} two seconds of data, recorded
on 14 November 2006 at h 23:30, are shown, as an example. The
amplitude values of the four channels, separately displayed, are
in V (the ADC input range was between -2V and +2V). A biological
sound is shown in figure \ref{fig:fig5_long_click}: the {\it
click} produced by a sperm whale (a signal emitted for
echo-location) and its reflection on the sea surface. A software
notch filter ($f=50$ Hz, $\Delta f = 50/35$ Hz, -10 dB, the same
for all channels) is applied to cut off the 50 Hz noise picked up
from the power system. The highest spectral components of sea
noise appear at $f<1$ kHz, thus they are filtered out in channels
H2 and H4. The electrical signal amplitude corresponding to the
click, recorded by H1,H2 and H3 is roughly the same, the signal in
H4 is about 5 dB smaller, as expected.

In order to determine the spectral Sound Pressure Density (SPD) of
sea noise, the Power Spectral Density (PSD) of the signal is
calculated per each recorded file (5' recording = 300$\cdot f_s$
samples) :

\begin{equation}
PSD(f)= \frac{|X_{N_{DFT}}|^2}{f_s \cdot L} \label{eq:PSD}
\end{equation}

where $f_s$ is the sampling frequency (96 kHz), $L$ is the time
length of the signal (in units of seconds), and $X$ is the
$N^{th}$ component of the Discrete Fourier Transform (DFT)
corresponding to the frequency $f$. The file is divided into
blocks of 2048 samples, weighted using an Hanning window and an
overlap of 50$\%$ (i.e. a 1024 samples shift). The 2048 points
Discrete Fourier Transform  ($\Delta f \simeq47$ Hz) is then
calculated using the Fast Fourier Transform algorithm implemented
on Matlab. Eventually we calculated the average, minimum and
maximum values and the $30^{\hbox {th}}$, $50^{\hbox {th}}$,
$90^{\hbox {th}}$ and $95^{\hbox {th}}$ percentile of the
distribution of the obtained PSDs.

The analysis presented in this paper was carried out using only
the data sample recorded with H3, from May to December 2005 and
from July to November 2006 ($\sim$ 6400 files). Other data are not
included in this paper, because they were taken using 16 bits
recording software, so they are not homogeneous with the rest of
the sample.

\subsection{Determination of the detector electronic noise floor}

\begin{figure}[h]
\centerline{\includegraphics[width=14cm]{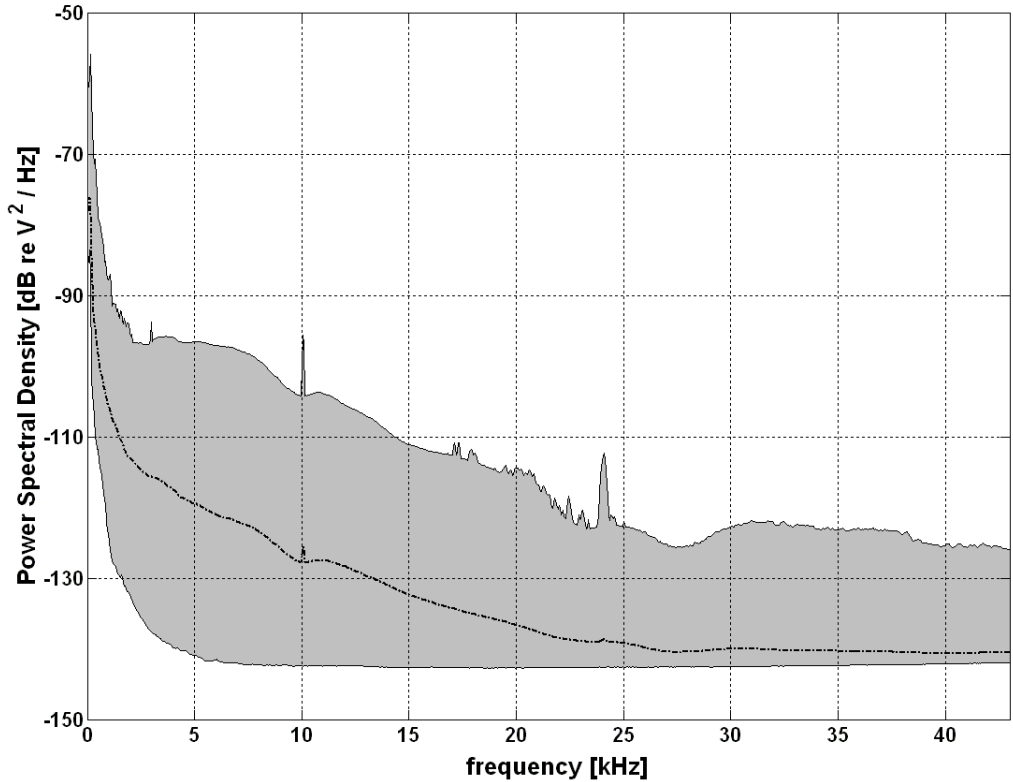}}
\caption{\label{fig:fig6_PSD_alldata} The grey area represents the
upper and lower limits of average PSDs calculated for the
$\sim6400$ files of data (the file time duration is 5') recoded
using channel H3. The black curve is the average calculated over
all PSDs.}
\end{figure}

Figure \ref{fig:fig6_PSD_alldata} presents the limits of
variations of the average PSD distributions (grey area) obtained
analysing $\sim$6400 files recorded by the hydrophones H3. Data
for $f>43$ kHz (0.45 $f_s$) are not shown. The plot shows large
variations in recorded signal amplitude, mainly for $f<$20 kHz,
and a baseline that represents the RMS power of the electronic
noise of our detector. This is a {\it white} noise\footnote{White
noise is a random signal with a flat power spectral density.} for
$f>5$ kHz recorded when the contribution of acoustic sea noise was
very low. As shown later, it is due to the self noise of the
hydrophone and the preamplifier, being the power of the ADC noise
negligible (few nV$^2$/Hz). The black dash-dotted line in figure
\ref{fig:fig6_PSD_alldata} represents the average of PSDs over the
whole data sample.

\begin{figure}[h]
\centerline{\includegraphics[width=14cm]{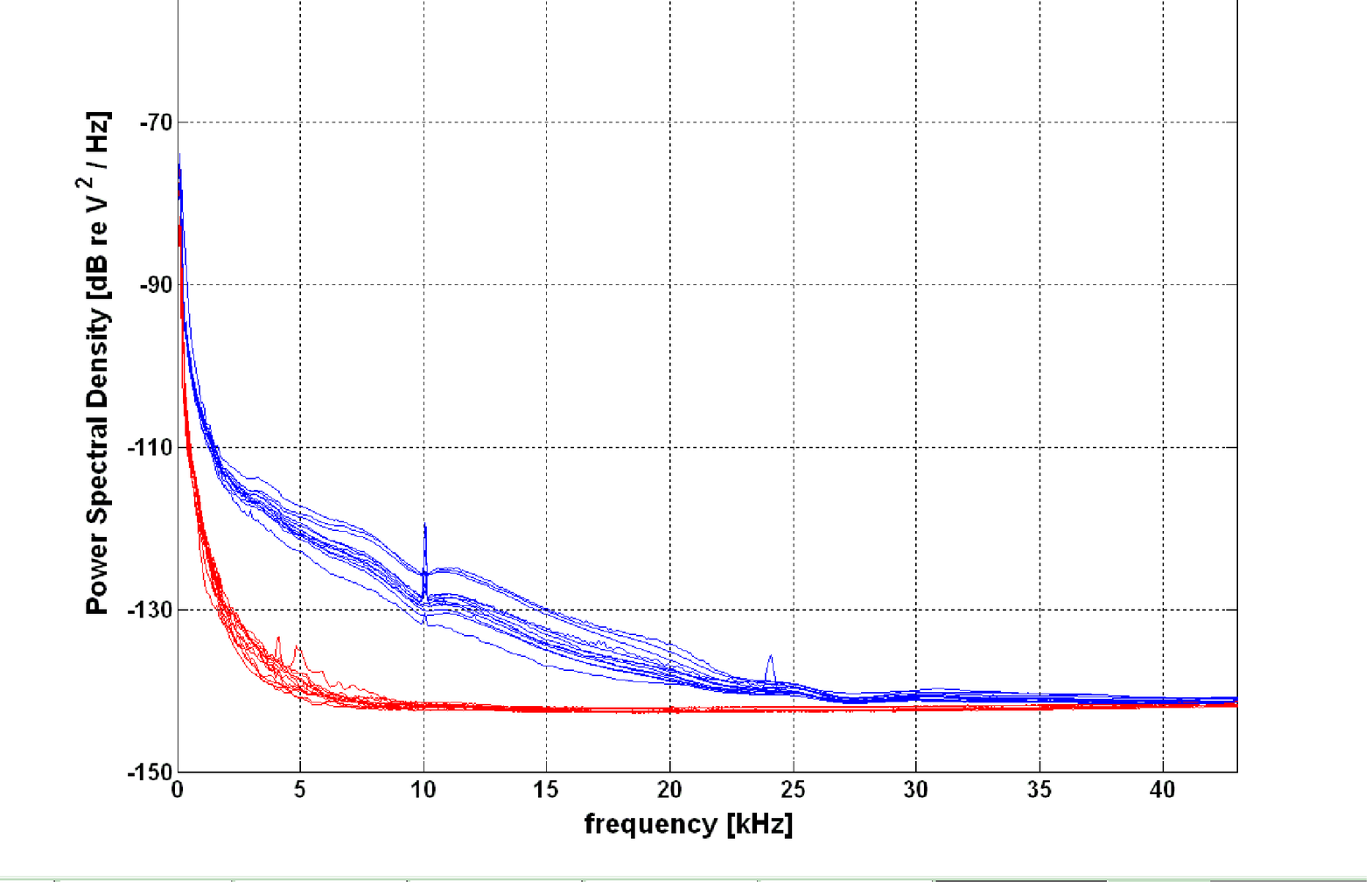}}
\caption{\label{fig:fig7_PSD_allmonths} Averages (blue lines) and
minima (red lines) of Power Spectral Density calculated, for each
month, using channel H3 data. The minima curves are nearly
superimposed: differences appears only at $f<5$ kHz, in months
showing a high acoustic background.}
\end{figure}

The average (in blue) and the minima (in red) of average PSD
curves calculated for different months with H3 are shown in figure
\ref{fig:fig7_PSD_allmonths}. While the average curves change for
different months, the minimum ones are very similar and almost
independent on the frequency, for $f>$5 kHz. This behaviour
indicates that minima are related to the electronics noise of the
detector (hydrophone coupled to the preamplifiers).

\begin{figure}[h]
\centerline{\includegraphics[width=14cm]{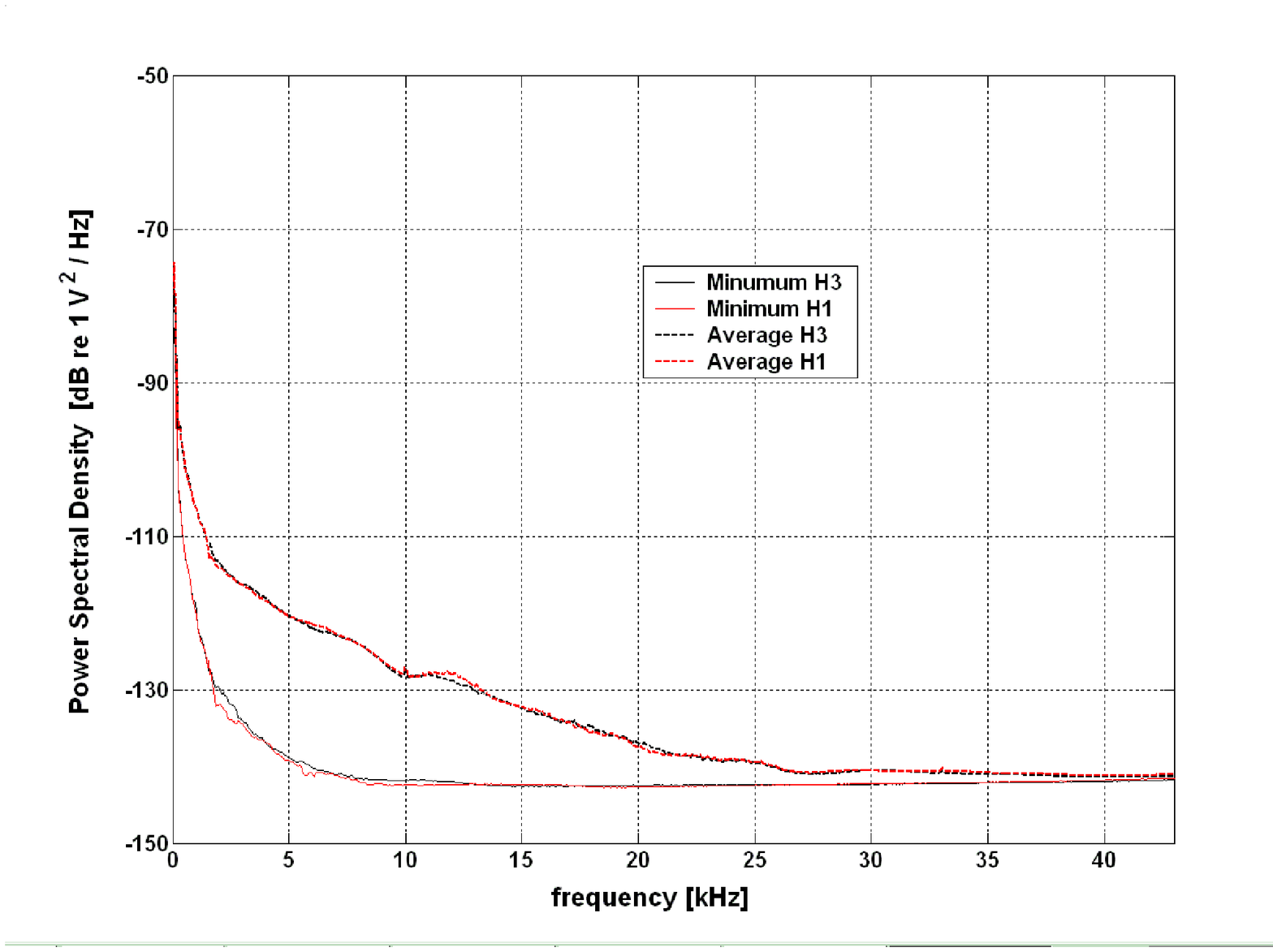}}
\caption{\label{fig:fig8_PSD_H1H3_Aug05} Average (dashed line) and
minimum (solid line) PSD curves calculated for about $740$ files
of data recoded during August 2005 with H3 (red) and H1 (black).}
\end{figure}

The same results are observed in all the channels: figure
\ref{fig:fig8_PSD_H1H3_Aug05} shows the minima (solid line) and
average (dashed line) curves calculated using the data recorded in
August 2005 for H1 (black) and H3 (red) respectively.

\begin{figure}[h]
\centerline{\includegraphics[width=14cm]{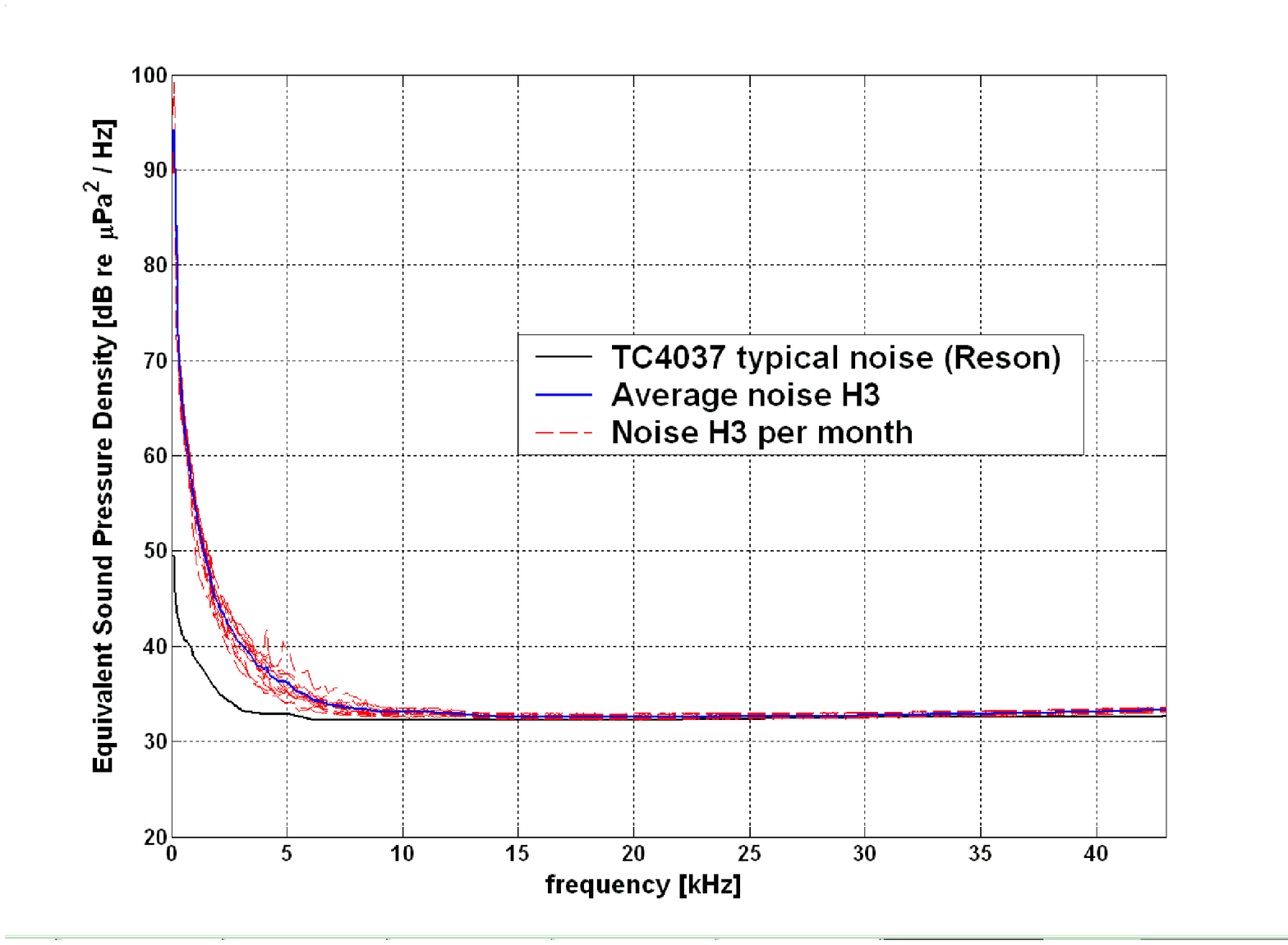}}
\caption{\label{fig:fig9_SPD_electronicsnoise_allmonths_uPa}
Equivalent sound pressure density of PSD minima calculated for
each month (red dashed lines) and their average value (blue thick
line). The equivalent noise curve provided by {\it RESON} for a
typical TC4037 hydrophone is shown for comparison (black curve).
The value of the minimum curve at $f>5 $kHz ($\simeq33\pm0.3$ db
re 1 $\mu$Pa$^2$/Hz) corresponds to the RMS of equivalent noise
power for a typical {\it RESON} TC4037 hydrophone and preamp
acquisition system.}
\end{figure}

In order to demonstrate the correlation between PSD (Power
Spectral Density) minima and electronic noise, the {\it
equivalent} sound pressure density of the PSD minima curves was
calculated, as shown in figure \ref{fig:fig7_PSD_allmonths}. The
SPD (Sound Pressure Density) curves, shown in figure
\ref{fig:fig9_SPD_electronicsnoise_allmonths_uPa} were obtained
multiplying PSD minima times the squared average sensitivity of
channel H3 (-195 + 20 dB re 1 V/$\mu$ Pa), assumed flat in
frequency. For $f>$5 the equivalent sound pressure density of PSD
minima is $\simeq 33 \pm$ 0.3 dB re $\mu$Pa$^2$/Hz. In this range
of frequencies the curves correspond in value and shape to the
power of the self noise estimated by the manufacturer for a
typical {\it RESON} TC4037 hydrophone and preamplifier set-up for
$f>$5 kHz\footnote{The {\it RESON} TC4042 is derived from TC4037
and used for larger depth applications} \cite{Reson}. At lower
frequencies this white electronics noise adds up with the noise
induced by the power supply and with the acoustic background (not
negligible at frequencies $\leq$ 1 kHz).

Since the electric signal produced by acoustic sea noise sums
incoherently with the electronics noise, the sound pressure
density of sea noise was recovered subtracting the average power
spectral density curve of noise from the PSD of the signal.

We also took the standard deviation of the power spectral density
minima curves distribution versus the month as a reference curve
(for each frequency) to indicate the systematic error in the
measurement of sound pressure density.

\subsection{First results}

\begin{figure}[h]
\centerline{\includegraphics[width=14cm]{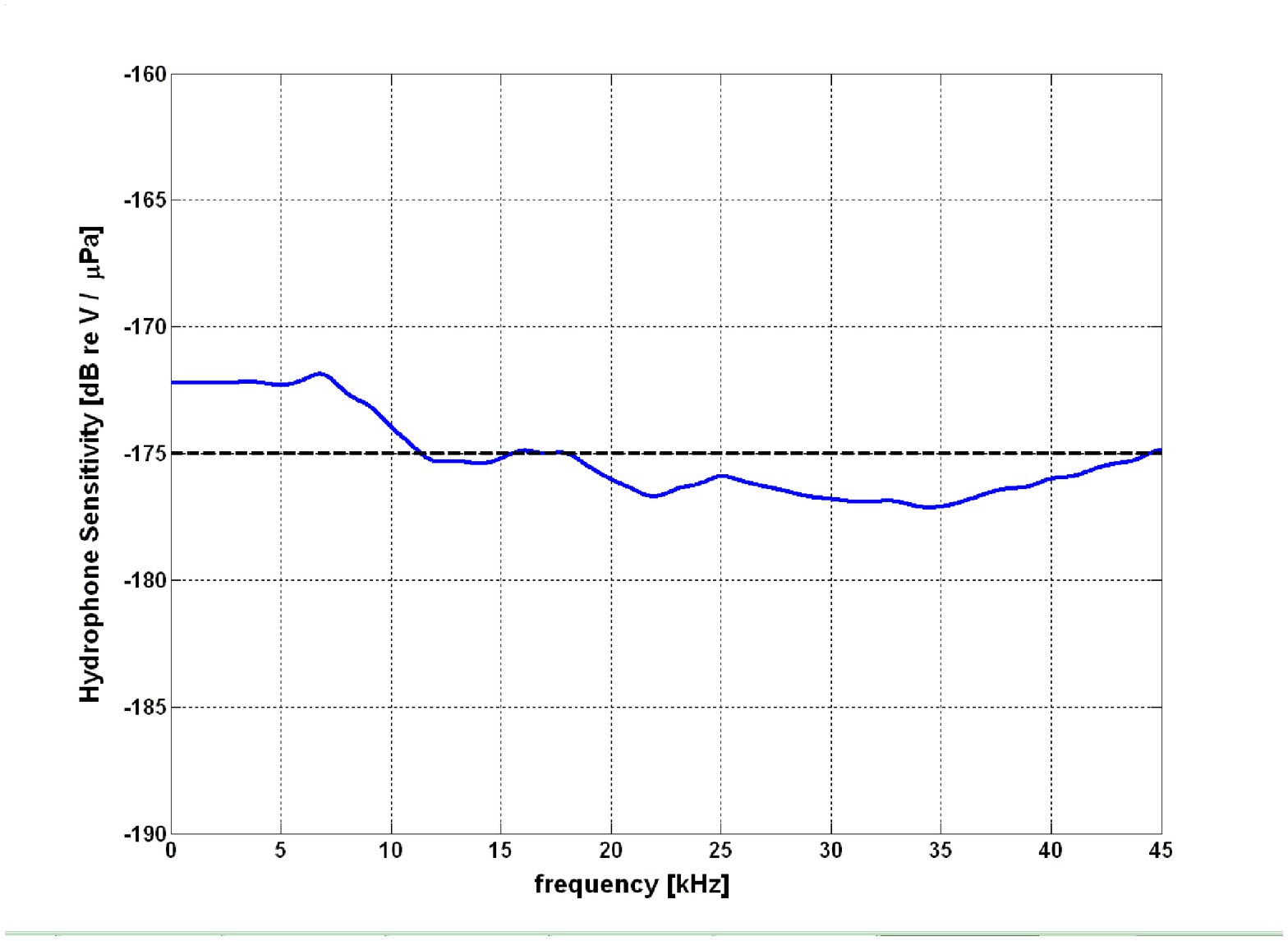}}
\caption{\label{fig:fig10_ResonSensitivity} Hydrophone sensitivity
(acoustic to electrical transduction factor) curve in units of dB
re V/$\mu$Pa as a function of frequency provided by the hydrophone
manufacturer for the hydrophone mounted in channel H3, the
premplifier gain is 20 dB. The average value of -175 dB used to
estimate the equivalent sound pressure density of electronic noise
is shown in black, as a reference.}
\end{figure}

Once the power spectral density of the electronic noise was
determined, the sound pressure density of environmental acoustic
noise was calculated taking into account the hydrophone
sensitivity curve given by the manufacturer, shown in figure
\ref{fig:fig10_ResonSensitivity}.

\begin{figure}[h]
\centerline{\includegraphics[width=14cm]{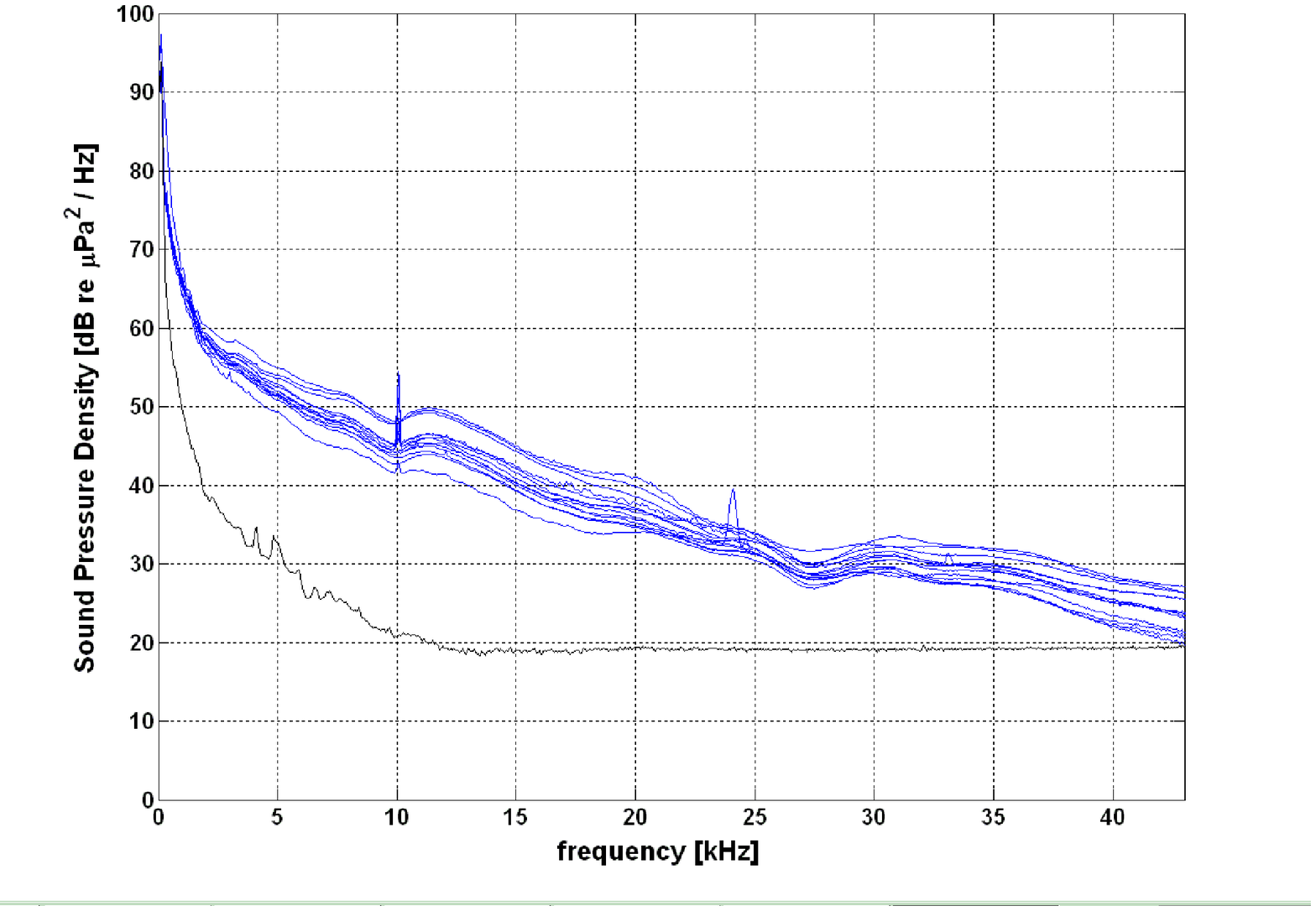}}
\caption{\label{fig:fig11_SPD_allmonths_uPa} Sound pressure
density curves of average sea acoustic noise as a function of
month (blue). The black curve indicates the systematic error in
the measurement due to the uncertainty on the electronic noise
power spectrum.}
\end{figure}

\begin{figure}[h]
\centerline{\includegraphics[width=14cm]{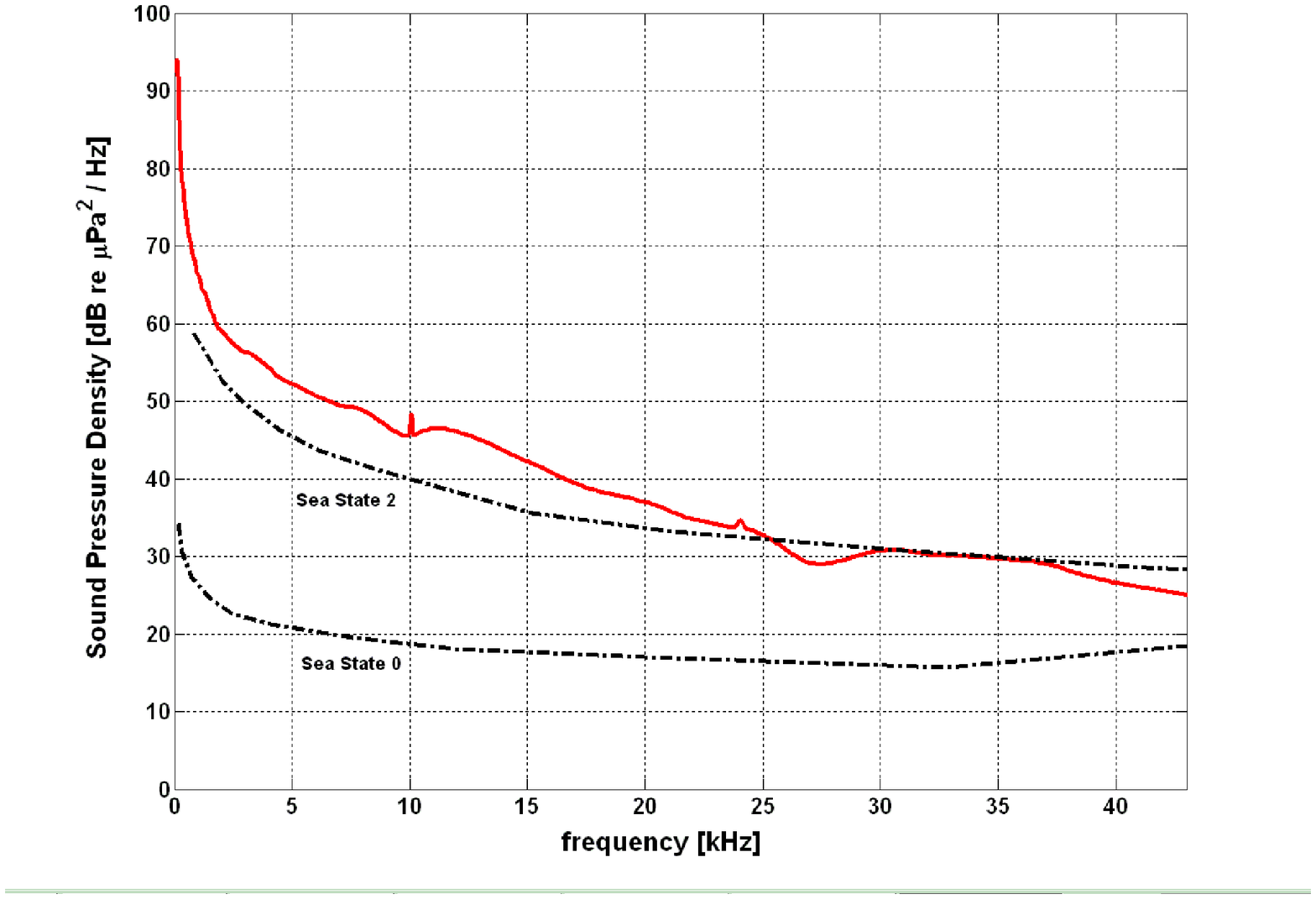}}
\caption{\label{fig:fig12_SPD_year_uPa} The blue solid line
indicates the average sound pressure density of sea noise recorded
with \onde channel H3 from May 2005 to November 2006. The dashed
blue lines indicate the systematic error on the average curve due
to uncertainty of the electronic noise power spectrum. The black
curves indicates respectively the expected SPD of the sea in
conditions of Sea State 0 and Sea State 2 \cite{Urick}.}
\end{figure}

In Figure \ref{fig:fig11_SPD_allmonths_uPa} is shown the sound
pressure density of the acoustic noise in deep sea calculated for
each month (blue curves), compared with the statistical error
curve (black) as defined in the previous section. In Figure
\ref{fig:fig12_SPD_year_uPa} is shown the average curve of sea
acoustic noise SPD calculated over the whole data sample (blue
line). The blue dashed curves take into account the error on the
electronic noise power determination. The black dashed curves
plotted in the same figure represent the sea noise sound pressure
density expected in conditions of Sea State Zero (SS0) and Sea
State Two (SS2) as defined by Urick \cite{Urick}, i.e. the SPD of
sea noise in conditions of absence of sea surface agitation (SS0),
or low surface agitation (SS2) and absence of identifiable
acoustic sources. A notable result for future underwater acoustic
neutrino experiments is that the average acoustic sea noise in the
band [$20\div43$ kHz] amounts to $5.4\pm2.2_{stat}\pm0.3_{syst}$
mPa RMS (the systematic error is due to the uncertainty on the
electronic noise power). This value is comparable to the estimated
acoustic signature of a 10$^{20}$ eV neutrino interacting at 1 km
distance from the detector (see reference
\cite{Vandenbroucke2005}).

\section{Capability of acoustic sources tracking }

Another characteristic of the station is the possibility to
reconstruct the direction of detected acoustic sources. This is
possible, in principle, since the antenna has a tetrahedral shape
and the hydrophone signals are synchronised and phased offline.
Nevertheless the hydrophone relative distances are about 1 m,
therefore for distant sources (i.e. in far field condition) \onde
can only recover the angular direction of the source in the
approximation of plane wavefront.

\begin{figure}[h]
\centerline{\includegraphics[width=14cm]{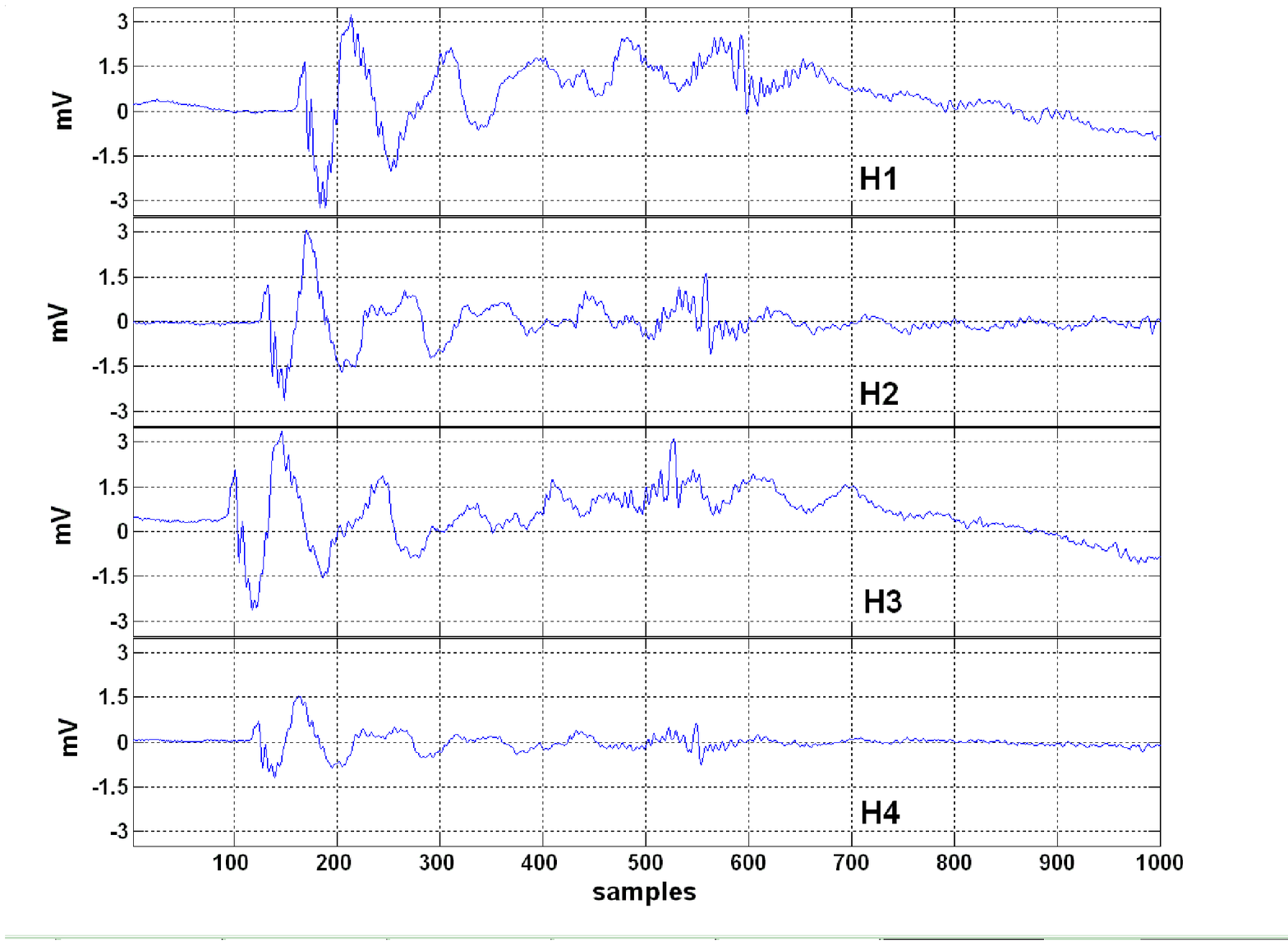}}
\caption{\label{fig:fig13_Click_TDOA} Click of a sperm whale
recorded by the four hydrophones. The amplitude scale is in mV.
Time Differences of Arrival (TDoA) between hydrophones are used to
recover the whale angular position.}
\end{figure}

As an example, in figure \ref{fig:fig13_Click_TDOA} is shown the
sperm whale {\it click} signal, quoted in figure
\ref{fig:fig5_long_click}, expanded in a different time-scale of
1000 samples (1/96 seconds). The signal is recorded first by H3
(installed on the frame apex) then by other hydrophones. Taking
into account phase correction, due to phase shift introduced by
the $f>1$ kHz hardware filter mounted on channels H2 and H4, the
time delay between H3 and the other channels was recovered using a
correlation function. Since the absolute position of hydrophones
in the frame and the absolute orientation of the frame with
respect to the sea bottom are known, the source angular direction
was calculated minimizing the functional
$G$($\vartheta$,$\varphi$), defined as:

\begin{equation}
\label{eq:trackingfunctional}
 G(\vartheta,\varphi)= \sum_{i=H1,H2,H4}[
TDOA^M_{i,H3}(\vartheta,\varphi) -
TDOA^E_{i,H3}(\vartheta,\varphi)]^2
\end{equation}

where $TDOA^M$ is the measured time delay of the signal recorded
by H3 with respect to the other hydrophones and $TDOA^E$ is the
expected time delay between H3 and H1, H2, H4 for signal coming
from a source located in the direction ($\vartheta$, $\varphi$).
The results indicate that, in this case, the sperm whale was
diving almost perpendicularly to the station ($\vartheta \simeq
87^{\circ}$) with a bearing of about 320$^{\circ}$ with respect to
North.

The estimate of angular direction reconstruction is affected by
systematic errors due to the uncertainties in relative hydrophones
distances (about 1 cm) and uncertainty in absolute frame
orientation (see section \ref{sec:electronics}). Preliminary
results carried out with a known source operated from the sea
surface indicate that this error is about $5^{\circ}$ both in
$\vartheta$ and $\varphi$, assuming sound propagation in a
homogeneous medium.

\section{Interdisciplinary activities}

The \onde experiment other than providing long term data on the
underwater noise, also provides a unique opportunity to study the
acoustic emissions of marine mammals living in the area or
transiting during their movements within the Mediterranean basin.
The most notable result was about the sperm whale presence and
transits in the area \cite{Science2007}. Several biological
sounds, unknown sounds and many man-made noises (ship and fishboat
noise, sonars, echosounders, airguns, and explosions) have been
recorded and archived for reference. The collection of acoustic
events collected so far represents a reference library to be used
for discriminating/separating known sources from potential
candidates of neutrino signatures in future larger dedicated
arrays. Concerning whale detection, the most common sounds
recorded are {\it clicks} produced by sperm whales arranged in
regular sequences (inter-click interval in the range 0.5 s to 2
s), or in special patterned sequences (chirrups, codas, creaks)
\cite{Pavan2000}. According to other studies in the Mediterranean
Sea, sperm whales may dive to more than 1000 meters depth, but
normally travel at 800$\div$1000 meters depth. Their source level
is typically $200\div220$ dB re 1 $\mu$Pa at 1 m on axis; the
loudest clicks were received with sound pressure levels up to 170
dB re 1 $\mu$Pa \cite{Pavan2000}. Clicks were often recorded with
high signal to noise ratio (SNR). Data from \onde indicate a
presence of sperm whales more consistent and frequent than
previously believed. Although the transiting of sperm whales is
known since the end of the XIX century \cite{Bolognari1949}, only
a small few literature is available for the area. IFAW
\cite{IFAW2003} reports a low sperm whale density in the Ionian
basin with an average encounter of 5.8 whale groups for 1000 km of
transect. With the \onde station, in year 2005, sperm whales have
been detected in 117 of the 231 (50.6\%) recorded days. The
analysis of 2006 data is still in progress but seems to confirm
the 2005 data \cite{Pavan2008}.

\section{Conclusions}

The \onde station successfully operated at the NEMO Test Site at
2000 m depth, 25 km offshore Catania (Sicily) from January 2005 to
November 2006. Mechanical, electronics and data transmission and
acquisition systems, designed and realised by INFN and CIBRA,
demonstrated high reliability and fulfilled electronic noise
design constraints. The station permitted for the first time a
long term characterization of deep sea noise in the Mediterranean
Sea in a large bandwidth ($0\div43$ kHz), with optimal signal
resolution. The electronics noise power spectrum  of the detector
was understood and evaluated to be $33 \pm 0.3$ dB re 1
$\mu$Pa$^2$/Hz above 5 kHz. It was also demonstrated that \onde is
capable to measure the sea noise Sound Pressure Density at the
reference level of Sea State Zero.

Data analysis is presently addressed to characterize the
underwater noise power level and its variations as a function of
time. The analysis carried out so far shows that the average Sound
Pressure Density of sea noise (over 13 months between May 2005 and
November 2006) agrees with equivalent Sound Pressure Density of
Sea State 2 for $f>25$ kHz. Larger values are recorded at lower
frequencies due to better propagation of lower frequencies sound
from surface to sea bottom and to man made noise (mainly ship
traffic). The identification of different noise sources is on
going.

The station also permits to localise noise sources using Time
difference of Arrival  of sound on the 4 hydrophones, in
conditions of plane wave approximation for far field sources. The
estimated error in the determination of the angular direction is
about 5 degrees in azimuth and zenith angles.

The recorded data set was extensively used for interdisciplinary
studies mainly addressed to search for cetaceans in the region;
results indicated an unexpectedly large number of detections
compared to previous studies.

\section{Acknowledgements}

The authors are grateful to L. Gualdesi and S. Buogo for help and
suggestions, to L. Dedenko, S. Danaher, L. Thompson and M. Ardid
for useful discussions. The authors also deeply thank the
electronics workshop (head C. Cal\'i) and the mechanical workshop
(head B. Trovato) of LNS-INFN for their support in the
construction of the experiment.

\end{document}